\documentclass[%
 aip,
 amsmath,amssymb,
 reprint,%
]{revtex4-1}

\usepackage{graphicx}
\usepackage{dcolumn}
\usepackage{bm}
\usepackage{wrapfig}

\usepackage[utf8]{inputenc}
\usepackage[T1]{fontenc}
\usepackage{mathptmx}
\usepackage{etoolbox}
\usepackage{comment}

\usepackage{array}
\usepackage{tabularx}

\usepackage{tabularx}
\usepackage{amsmath}
\usepackage{subcaption}

\begin{document}

\preprint{arXiv}

\title{Reversible single-pulse laser-induced phase change of Sb$_2$S$_3$ thin films: multi-physics modeling and experimental demonstrations} 

\author{Capucine Laprais}%

\author{Clément Zrounba}%
\author{Julien Bouvier}%
\affiliation{CNRS, Ecole Centrale de Lyon, INSA Lyon, Universite Claude Bernard Lyon 1, CPE Lyon, INL, UMR5270, 69130 Ecully, France}%
\author{Nicholas Blanchard}%
\affiliation{Universite Claude Bernard Lyon 1, CNRS, Institut Lumière Matière, UMR5306, F-69100, Villeurbanne, France}%
\author{Matthieu Bugnet}%
\affiliation{CNRS, INSA Lyon, Universite Claude Bernard Lyon 1, MATEIS, UMR 5510, 69621 Villeurbanne, France}%
\author{Yael Guti{\'e}rrez}%
\affiliation{Departamento de Física Aplicada, Universidad de Cantabria, Avenida de los Castros, s/n, Santander, 39005 Spain}%
\author{Saul Vazquez-Miranda}%
\affiliation{ELI Beamlines Facility, The Extreme Light Infrastructure ERIC, Za Radnicí 835, 25241 Dolní Břežany, Czech Republic}%
\author{Shirly Espinoza}%
\affiliation{ELI Beamlines Facility, The Extreme Light Infrastructure ERIC, Za Radnicí 835, 25241 Dolní Břežany, Czech Republic}%
\author{Peter Thiesen}%
\author{Romain Bourrellier}%
\affiliation{Park Systems France, 21 rue Jean Rostand, 91400 Orsay}%
\author{Aziz Benamrouche}%
\author{Nicolas Baboux}%
\author{Guillaume Saint-Girons}%
\author{Lotfi Berguiga}%
\author{Sébastien Cueff}%
\affiliation{CNRS, Ecole Centrale de Lyon, INSA Lyon, Universite Claude Bernard Lyon 1, CPE Lyon, INL, UMR5270, 69130 Ecully, France}%

\email{capucine.laprais@ec-lyon.fr}

\date{\today}

\begin{abstract}
Phase change materials (PCMs) have gained a tremendous interest as a means to actively tune nanophotonic devices through the large optical modulation produced by their amorphous to crystalline reversible transition. Recently, materials such as Sb$_2$S$_3$ emerged as particularly promising low loss PCMs, with both large refractive index modulations and transparency in the visible and NIR. Controlling the local and reversible phase transition in this material is of major importance for future applications, and an appealing method to do so is to exploit pulsed lasers. Yet, the physics and limits involved in the optical switching of Sb$_2$S$_3$ are not yet well understood. Here, we investigate the reversible laser-induced phase transition of Sb$_2$S$_3$, focusing specifically on the mechanisms that drive the optically induced amorphization, with multi-physics considerations including the optical and thermal properties of the PCM and its environment. We theoretically and experimentally determine the laser energy threshold for reversibly changing the phase of the PCM, not only between fully amorphous and crystalline states but also between partially recrystallized states. We then reveal the non-negligible impact of the material's polycrystallinity and anisotropy on the power thresholds for optical switching. Finally, we address the challenges related to laser amorphization of thick Sb$_2$S$_3$ layers, as well as strategies to overcome them. These results enable a qualitative and quantitative understanding of the physics behind the optically-induced reversible change of phase in Sb$_2$S$_3$ layers. \\
\end{abstract}

\maketitle

\section{\label{sec:level1}Introduction}

Recent advances in nanophotonics have opened up unprecedented opportunities for controlling light matter interactions at the nanoscale. Yet, as of today, most devices are passive, and hence, limited to one dedicated functionality set at the fabrication stage. Implementing tunable and reconfigurable capabilities on devices appear now as a necessity to unlock the full potential of nanophotonics. For that matter, functional materials have recently gained attention for tunable nanophotonics as they can allow a dynamic control of light without any moving parts \cite{cui2019tunableMSreview}. 

In this regard, phase change materials (PCMs) are highly promising, when integrated with waveguides and metasurfaces, for applications in telecommunications, displays, sensing or neural network \cite{Cao2020, wuttig2017phase}. Their particularity is the possibility to rapidly and reversibly change their state from amorphous to crystalline. This phase transition is non-volatile and induces a much larger change in refractive index than other tuning mechanisms \cite{abdollahramezani2020tunable,mikheeva2022space}. Most of the works on PCMs focus on standard PCMs such as GST or GeTe, used as switches between the amorphous and crystalline state. Although these PCMs have been successfully used as memory elements on waveguides \cite{rios2015integrated}, as artifical synapses and neurons \cite{bruckerhoff2023event,cheng2017chip,feldmann2019all}, for tunable absorbers \cite{kats2012ultra,chen2015tunable,cueff2021}, beam steerers \cite{yin2017beamSterring,de2018nonvolatile} or color filters \cite{Hosseini2014}, they are often too lossy for specific applications in the visible and near infrared ranges \cite{wang2023ReviewPCM}. For this reason, a new class of emerging PCMs, so-called “low loss PCMs”, have been the subject of increasing interest. Sb$_2$S$_3$ is one of these PCMs, it has recently been introduced to the photonics community by Dong et al.\cite{dong2019Sb2S3}  for its exciting optical properties: beyond 780 nm it presents a negligible absorption in both phases while its refractive index contrast remains very large ($\Delta$n=1) \cite{dong2019Sb2S3,Delaney2020,bieganski2024}. However, being an emerging PCM, Sb$_2$S$_3$ has not yet reached the maturity of GST, and many of its functional aspects need to be understood and optimized, especially its switching performances (e.g. switching time, cyclability) \cite{wang2023ReviewPCM}. Additionally, recent years have witnessed a paradigm shift towards prioritizing multi-level phase transition rather than using PCMs as simple binary switches. While controlling the partial recrystallization of unpatterned Sb$_2$S$_3$ thin films has been demonstrated \cite{programPCM}, only a few studies reported the control of the reversible partial phase transition at the device level \cite{wei2024monolithic,chen2023,yang2023non}. Accordingly, developing approaches for locally controlling the phase transition of Sb$_2$S$_3$ is essential and demands a comprehensive understanding of the involved phenomena. \\ 
Among the possible methods for inducing the phase transition, the optical switch is very versatile as it allows a local and reversible phase change without additional fabrication steps such as the integration of micro-heaters. The physical principle of the switch is straightforward: the laser energy absorbed within the PCM generates heat, which subsequently provokes the change of phase once the crystallization or metling temperature is reached. However, optimizing the reversible change of phase requires dealing with a multi-physics interplay between laser absorption, heat transfer and phase change dynamics \cite{lawson2024MilionCycle}. The majority of studies on laser-induced phase transition in PCMs concentrate on image printing \cite{Liu2020,Wang2014GSTGreyScale,Hosseini2014} as well as writing devices such as gratings \cite{Hebler2021IST} or Fresnel zone-plates \cite{Wang2016,gao2024SbSOpticalswitchLambda}, without in-depth description of all the underlying mechanisms involved in the reversible optical switch. Consequently, our work aims at gaining insights in the physical mechanisms at play when switching optically Sb$_2$S$_3$ by relating multi-physics (thermal, optical and phase change) simulations to experimental results. 

Here, we focus on the reversible laser amorphization of Sb$_2$S$_3$  realized with a single-pulse laser, contrarily to most reports where the phase transition is generally performed using several pulses \cite{Gao2021,Liu2020,sun2017GeTeOptPC}. Specifically, we determine the energy range for the reversible phase change of Sb$_2$S$_3$ both experimentally and via simulations. Furthermore, by investigating the laser recrystallization dynamics, we achieved the partial phase transition of Sb$_2$S$_3$. We also report the impact of Sb$_2$S$_3$ anisotropy on the laser induced amorphization, and identify the limiting factors in optically programming thick Sb$_2$S$_3$ layers. Finally, we investigate the impact of the film thickness and, based on the simulations, propose a strategy to switch thick films, which is often reported challenging \cite{zhang2021myths,lawson2024MilionCycle}. 

\section{Laser amorphization principle and theory}
\begin{figure*}[!ht] 
\centering
   \includegraphics[width=0.9\linewidth]{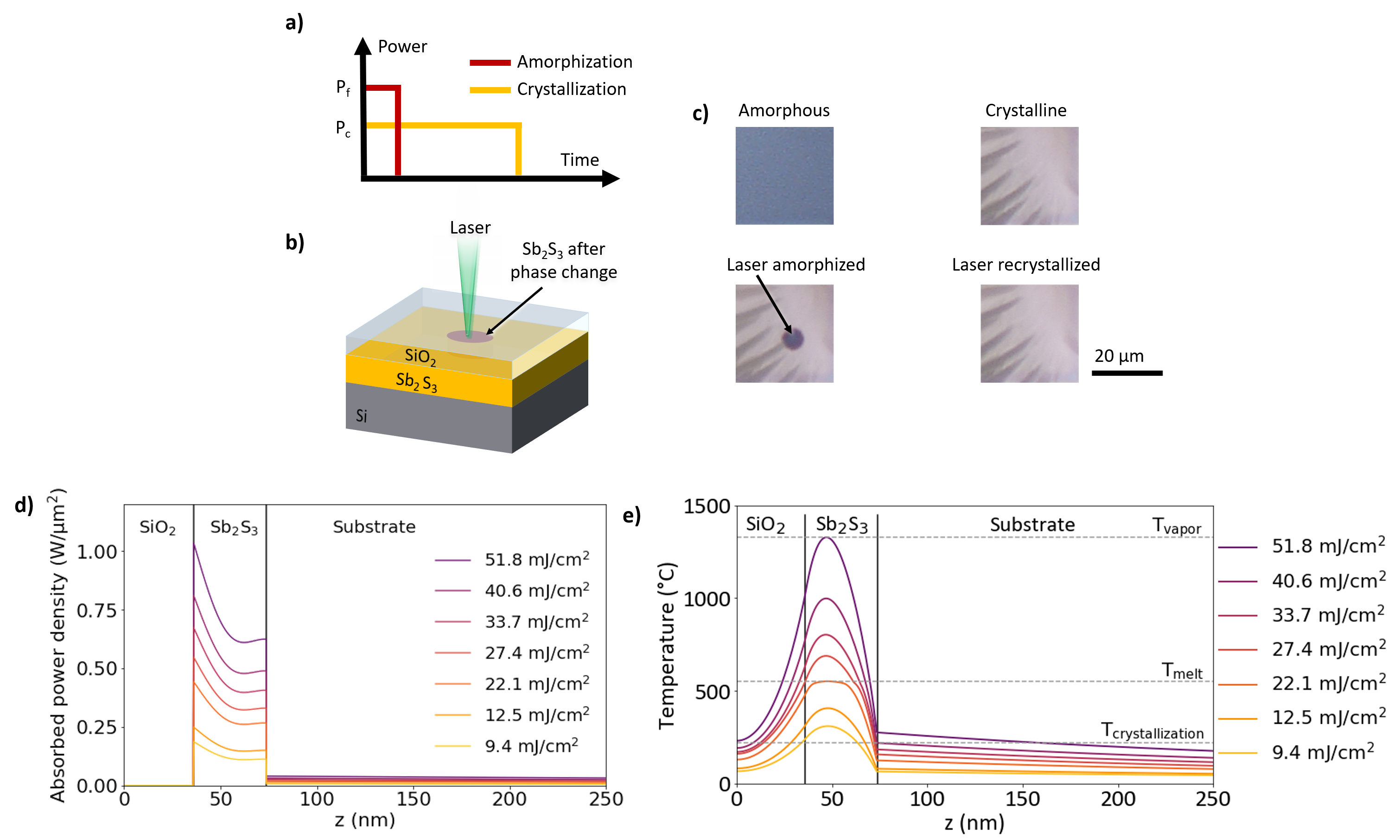}
  \caption{Optical switch principle  \textbf{a)} laser pulse profile for each phase transition, \textbf{b)} schematic of Sb$_2$S$_3$ optical switch principle, \textbf{c)} optical microscope image of the 42 nm-thick Sb$_2$S$_3$ sample in its different states, \textbf{d)} and \textbf{e)} simulated optical absorption and corresponding temperature profile at the end of the 500 ps pulse along the stack, for a 42 nm-thick Sb$_2$S$_3$ layer, plotted for several values of laser pulse fluence.}
 \label{Principle}
\end{figure*}

The phase transitions in PCM are temperature-time dependent transformations. Typically, amorphization is the result of a melt-quench process demanding high temperature (above the melting point) to melt the PCM and short heating time so that the PCM quickly cools down without recrystallizing. Conversely, crystallization necessitates longer heating time at moderately high temperatures (below the melting point) to let the crystals form and grow. 
To induce these phase changes optically, the material is heated by laser irradiation: the light is substantially absorbed in the film depending on the PCM's optical properties, which translates into a localized temperature elevation. For a given laser wavelength, the type of phase transition (amorphization or crystallization) is determined by the laser power and exposure time as shown in figure \ref{Principle} \textbf{
a)}. 

Here, we focus on the reversible laser amorphization of Sb$_2$S$_3$ thin films exposed to a 532 nm laser at normal incidence, as illustrated in figure \ref{Principle} \textbf{b)}. This laser energy (2.33 eV) is over the bandgap of the crystalline Sb$_2$S$_3$ (1.75 eV) and it is absorbed by the amorphous Sb$_2$S$_3$.
Examples of experimental demonstrations of the phase transition are provided in figure \ref{Principle} \textbf{c)}, representing an optical microscope top view of a 42 nm-thick Sb$_2$S$_3$ film in its two reference states (amorphous and crystallized - top images) and after laser exposure (bottom images).
The circular laser-amorphized region (bottom left image) appears with the same purple color as the amorphous reference and is completely erased after laser recrystallization (bottom right image).
To better understand the physical mechanisms involved in the laser-induced amorphization of Sb$_2$S$_3$, we simulate the laser heating using an in-house multiphysics model comibining optical and thermal simulations of the device.

\subsection{Optical simulation}

The light absorbed within the stack containing the Sb$_2$S$_3$ thin film determines the amount of energy available for the phase transition and depends on the optical absorption of Sb$_2$S$_3$ as well as on the multiple reflections within the different layers.
The optical behaviors of the stack (absorption, transmission, and reflection) under normal incidence are numerically calculated by employing the transfer matrix method \cite{mackay2022transfer}.
The simulated stack consists in a silicon substrate, a 42 nm-thick Sb$_2$S$_3$ layer and a 30 nm-thick SiO$_2$ capping layer.
The optical properties of the Sb$_2$S$_3$ and SiO$_2$ thin films (refractive index n and absorption coefficient k) used for the calculation are: $\mathrm{n_{Sb_2S_3  c}}=4.72 \pm 0.03$, $\mathrm{k_{Sb_2S_3  c}}=0.74 \pm 0.04$, $\mathrm{n_{SiO_2}}=1.47$ and $\mathrm{k_{SiO_2}}=0$.
These values were determined experimentally by previous ellipsometry measurements \cite{programPCM,bieganski2024}.
The laser wavelength is fixed at the experimental value of 532 nm.
Since the simulated profile is obtained by performing 1D simulations, the spot size is not directly included in the calculation.
Instead, the laser pulse fluence is used to calculate the expected absorbed power density in the stack from the integration of the Poynting vector.
The resulting absorbed power density profile, shown in figure \ref{Principle} \textbf{d)}, does not vary uniformly across the Sb$_2$S$_3$ layer: it decreases sharply in the first 25 nm of the PCM before holding steady and slightly increasing in the last nanometers.
This is an expected result from the stack geometry, which produces standing waves in the PCM layer.
Naturally, increasing the laser fluence also leads to higher absorbed power density.
The resulting absorption profile is then used for solving the heat equation and obtaining the corresponding temperature throughout the film, as explained in details in the following.

\subsection{Thermal simulation}

The temperature profile in the stack, illustrated in figure \ref{Principle} \textbf{e)}, is obtained by solving the 1D heat equation in the structure:
\begin{equation}
    \rho C_p \frac{\partial T}{\partial t} = \frac{\partial}{\partial z} \left ( \kappa  \frac{\partial T}{\partial z} \right ) + Q
\end{equation}
Where $\rho$ is the density, $C_p$ is the specific heat capacity at constant pressure, $T$ is the temperature, $t$ is the time, $z$ is the distance to the surface of the sample, $\kappa$ is the thermal conductivity and $Q$ is the absorbed power density calculated in the optical simulation as explained above.
For the calculation, the bottom temperature of the substrate is fixed at room temperature and the top surface is considered as thermally isolated ($\frac{\partial T}{\partial x}=0$). 
The lateral heat transfer is not taken into account in the 1D simulations, and the thermal boundary resistances from layer to layer are also ignored due to the lack of experimental data. 

\begin{figure}[h!] 
\centering
   \includegraphics[width=0.8\linewidth]{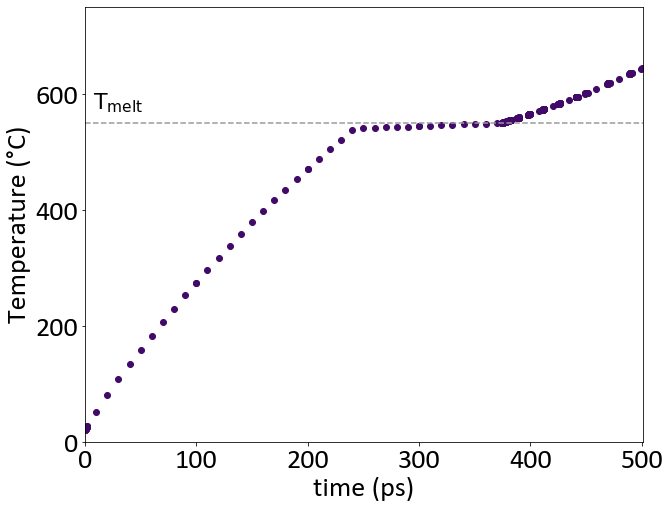}
  \caption{Maximum simulated temperature in the PCM film taken every 10 ps during the entire duration of the pulse, for a 27.4 mJ/cm$^2$ pulse fluence}
 \label{SimuEvolTemp}
\end{figure}

As the thermal simulation progresses, the PCM temperature is monitored (see figure \ref{SimuEvolTemp}), so as to interrupt the simulation as soon as the melting temperature is attained in any location within the PCM layer. 
When it does occur, the layer structure is updated to account for the newly melted region, for which amorphous material properties are assumed, and a new optical simulation is run.
This sequential multi-physics simulation continues along the duration of the single pulse.
In order to account for the latent heat of fusion, the specific heat of crystalline Sb$_2$S$_3$ is adjusted right below the melting point.
This predictably leads to a plateauing effect near $\mathrm{T_{melt}}$, as can be observed in figure \ref{SimuEvolTemp} illustrating the temperature rise in the PCM film throughout the duration of the pulse.

Figure \ref{Principle} \textbf{e)} shows examples of temperature profiles calculated across the depth of the 42 nm-thick Sb$_2$S$_3$ sample against laser fluence, for a fixed pulse duration of 500 ps (chosen to match our experimental conditions).
The calculated temperature profile within the stack takes on a bell shape that is markedly different from the optical absorbed power density profile (see Fig. \ref{Principle} d) and e) ), showing the non-negligible impact of heat transfer.
Note that the peak temperature is located closer to the capping than the substrate. This is because most of the heat is evacuated through the highly thermally-conductive Si layer, whereas the thin SiO$_2$ capping layer acts as a temporary heat store since it is thermally insulated.
Consequently, Sb$_2$S$_3$ starts amorphizing at the location of the peak temperature and the amorphized region subsequently spreads across the PCM layer. The melting temperature is first reached 10 nm below the capping/Sb$_2$S$_3$ interface, for a laser fluence of at least 22 mJ/cm$^2$. The top surface of the PCM then starts to melt for a laser fluence of 27.4 mJ/cm$^2$.
By further increasing the laser power, a larger fraction of the material can be amorphized, but, for a laser fluence above 51.8 mJ/cm$^2$, the peak temperature reaches the evaporation point of Sb$_2$S$_3$. We consider this fluence (51.8 mJ/cm$^2$) as a threshold limit before damaging the PCM layer.
The fluence window for amorphizing Sb$_2$S$_3$ thus spans from 22 mJ/cm$^2$ to 51.8 mJ/cm$^2$. For a lower pulse fluence (between 22 mJ/cm$^2$ and 27.4 mJ/cm$^2$) the PCM layer has a crystalline-amorphous-crystalline structure, while a higher fluence (between 27.4 mJ/cm$^2$ and 51.8 mJ/cm$^2$) results in an amorphous-crystalline structure. 
For the specific case presented here, the film is never fully amorphized: at most, the amorphized depth is 34 nm when the maximum temperature reaches the evaporation point.
Not fully amorphizing the film could limit the applications of Sb$_2$S$_3$, hence the need for thorough experimental verification of these results. \\

\section{Experimental results}

We describe in the following the experimental demonstrations of reversible laser amorphization of a 42 nm-thick Sb$_2$S$_3$ film. The PCM is deposited amorphous on a silicon substrate by electron beam evaporation and capped with a SiO$_2$ layer (30 nm-thick) to prevent oxidation \cite{teocapping}. The amorphous as-deposited Sb$_2$S$_3$ layer is then thermally crystallized under ambient conditions at 280°C, before being laser amorphized. The reversibility of the laser amorphization is verified by recrystallizing the PCM as mentioned previously (see figure \ref{Principle} \textbf{c)}).

\subsection{Laser switching optical setup}
\begin{figure}[ht!] 
\centering
   \includegraphics[width=1\linewidth]{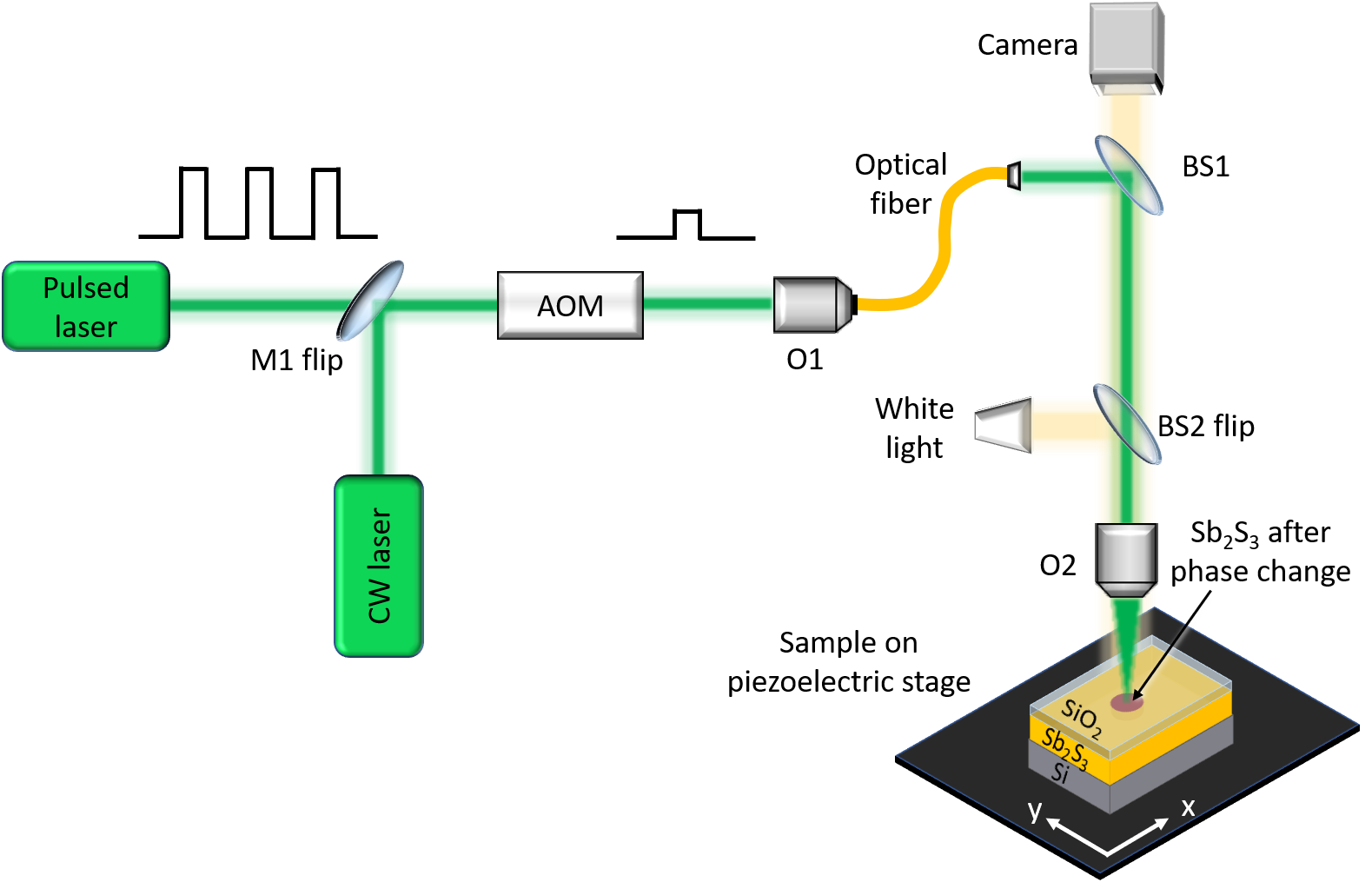}
  \caption{Optical set-up used for the reversible phase transition of the PCMs.}
 \label{Setup}
\end{figure}

 The optical set-up used for the laser-induced phase transition is illustrated in figure \ref{Setup}. Amorphization is realized with a laser with a fixed pulse duration of 500 ps, while recrystallization of the amorphized spots is performed with a continuous wave (CW) laser. The two lasers have the same wavelength: 532 nm, and are focused over a 10 µm diameter spot on the top surface of the sample. An acousto-optic modulator (AOM) controlled by a function generator and a LabView program enables: (i) selecting a single laser pulse for amorphization, (ii) defining the exposure time of the continuous laser for recrystallization, and (iii) tuning both laser powers. \\

\subsection{Laser amorphization}\label{Laser amorphization}

\begin{figure*}[ht!] 
\centering
   \includegraphics[width=0.7\linewidth]{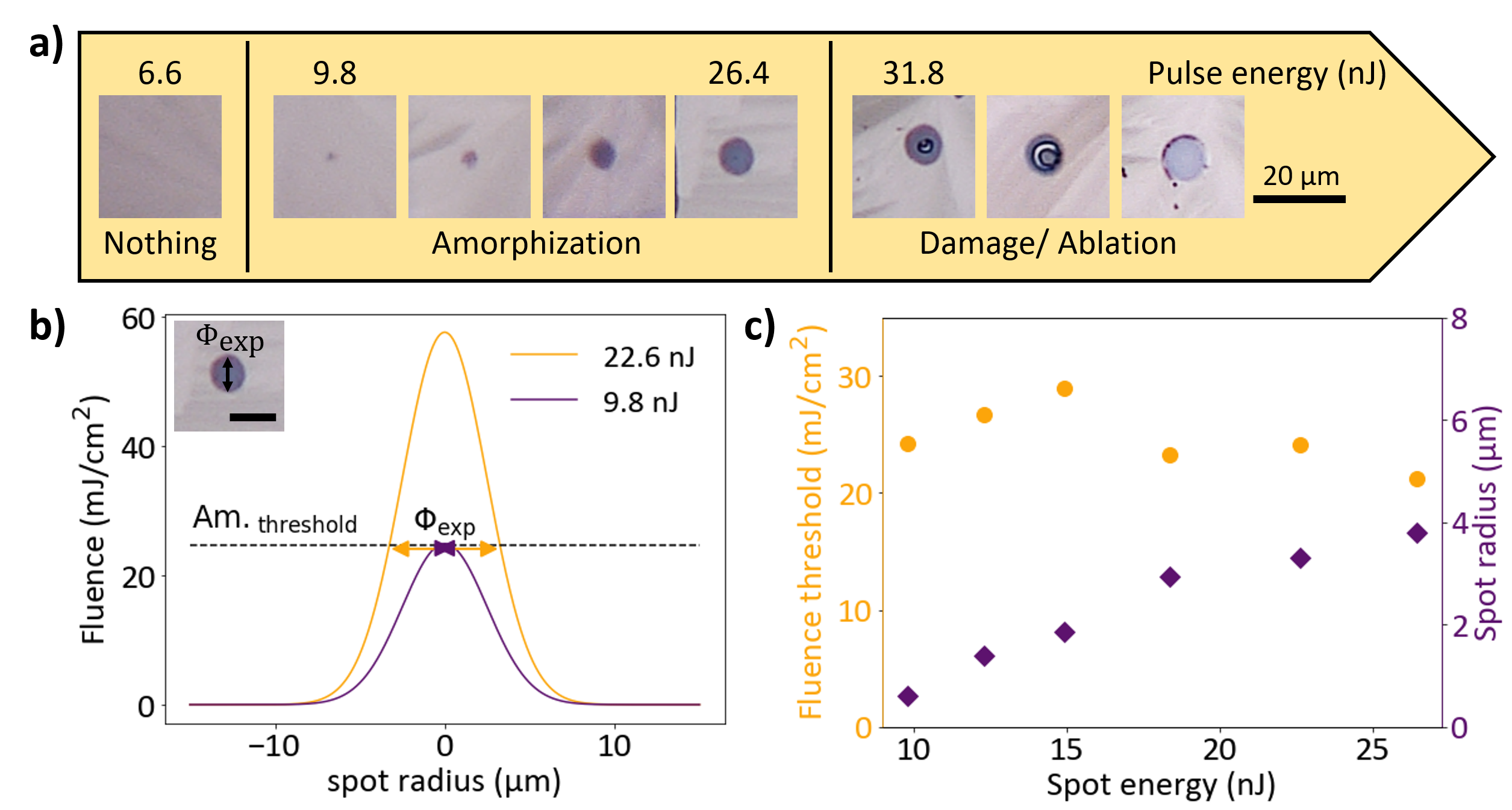}
  \caption{Determination of the minimum fluence threshold for amorphization with \textbf{a)} optical microscope images of the spot as a function of the pulse energy, \textbf{b)} example of the calculated Gaussian profile of the spot for two different pulse energies, and \textbf{c)} spot radius and fluence threshold as a function of the pulse energy.}
 \label{ExpAm}
\end{figure*}
To experimentally determine the minimum laser fluence for amorphizing the Sb$_2$S$_3$ sample, we sweep the laser power and first rely on colorimetric observations for identifying amorphized regions. Given the different optical properties of the two phases, amorphized spots should appear in different colors than their surroundings when observed under an optical microscope. The results are provided in figure \ref{ExpAm} \textbf{a)}, where the optical microscope images of the sample's top surface are displayed as a function of the pulse energy. For an energy between 9.8 nJ and 26.4 nJ, the purple color of the amorphized Sb$_2$S$_3$ is clearly distinguishable. Below 9.8 nJ, no change on the film surface compared to the crystalline reference is noticeable. Above 26.4 nJ, the center of the spot exhibits a dark color that cannot be recrystallized. We concluded that the surface of the sample starts to be damaged and is eventually ablated when the laser energy is further increased. This damage corresponds to a swelling of the film (see AFM characterization in the supplementary), and is likely due to the thermal expansion of the PCM under excessive heating. As the capping layer can no longer accommodate for the PCM's volume change it is irremediably pushed away, resulting in the film's ablation. For amorphizing Sb$_2$S$_3$, the laser energy thus should be kept between 9.8 nJ and 26.4 nJ. 

We also notice subtle changes within this amorphizaton window: the amorphized region expands and the spot color becomes more saturated as we increase the laser pulse energy. We tentatively ascribe this tendency to a partial amorphization for low energy spots, as predicted by the simulations. The spot size dependency to the pulse energy is attributed to the Gaussian profile of the beam: since the laser intensity is the highest in the center of the spot, this area would reach the melting point before the edges of the spot do. As a consequence, for the spots obtained with a low pulse energy, only the center of the beam is powerful enough to melt Sb$_2$S$_3$, while for high energies nearly all the surface of the beam provides enough power for amorphization, as illustrated in figure \ref{ExpAm} \textbf{b)}. We can extract the experimental amorphization fluence threshold by measuring the spots diameters as a function of the laser energy and calculating the corresponding intensity at the edge of the spot using the Gaussian beam formula: 
\begin{equation}
   F(r)=\frac{2E}{\pi w_0^2}exp(\frac{-2r^2}{w_0^2})  
\end{equation}
where $F$ is the laser fluence, $E$ is the measured pulse energy arriving on the sample, $w_0$ is the beam waist (here 10 µm) and $r$ is the distance to the center of the beam spot.
These results are given in figure \ref{ExpAm} \textbf{c)} for six different pulse energies. Using this method, the experimental amorphization threshold was found to be between 21.2 mJ/cm$^2$ and 28.9 mJ/cm$^2$, which is in good agreement with the simulations. The 7.7 mJ/cm$^2$ uncertainty can be attributed to measurement uncertainty of the laser power and spot diameter, as well as to the material's anisotropy as will be discussed further. The damage threshold is harder to estimate as the shape of the damaged area is more unpredictable. It should however lie between 67.3 mJ/cm$^2$ (maximum of the Gaussian for 26.4 nJ laser energy) and 79 mJ/cm$^2$ (maximum of the Gaussian for the 31.8 nJ spot). In this case, the damage threshold was under-estimated by the simulations.

\begin{figure}[ht!] 
\centering
   \includegraphics[width=1\linewidth]{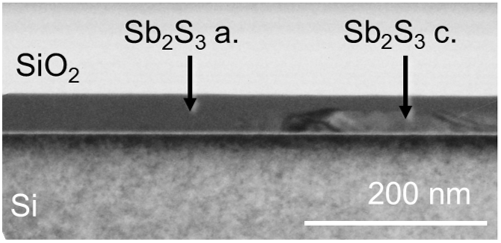}
  \caption{TEM cross section of an amorphized spot observed in bright field mode on the 42-nm-thick Sb$_2$S$_3$ layer.}
 \label{TEM42}
\end{figure}

We now want to investigate the depth of the amorphization spots. As mentioned before, the low power spots appear in lighter color compared to the amorphous reference, which suggests partial amorphization. Additionally, from the simulation, the maximum depth that could be amorphized is 34 nm. Consequently, although a simple color comparison allows to know if the phase transition occurred, the thickness of the change of phase remains imprecise. To gain insights on the amorphization depth, a TEM cross section of the spot was performed and the edge of the amorphized region is shown in Figure \ref{TEM42}. The observed spot was obtained with the highest possible laser power below the damage threshold to maximize the chance of amorphizing the sample down to the substrate. As displayed in Figure \ref{TEM42}), the TEM observation confirms that the Sb$_2$S$_3$ layer was completely amorphized in the center of the spot, contrary to the simulations prediction, with the amorphous region appearing homogeneous as a result of no diffraction contrast. The Gaussian shape of the beam is also visible at the edge of the spot and, as predicted by the simulations, the amorphization starts close to the surface and extends in the PCM's depth. This observation thus confirms the analysis presented above as well as the possibility to fully amorphize 42 nm of Sb$_2$S$_3$. It is also a clear demonstration that our simulations provide a good description of the physical mechanisms involved in the laser induced amorphization. The slight differences with the experiments are attributed to experimental uncertainties and simulation hypothesis. Indeed, only 1D simulations were performed and thus neither the Gaussian profile of the beam nor the heat dissipation on the sides (in x and y directions) were taken into account. These approximations probably induce discrepancies in the results.

\subsection{Laser recrystallisation}
\begin{figure}[!ht] 
    \centering
    \includegraphics[width=0.9\linewidth]{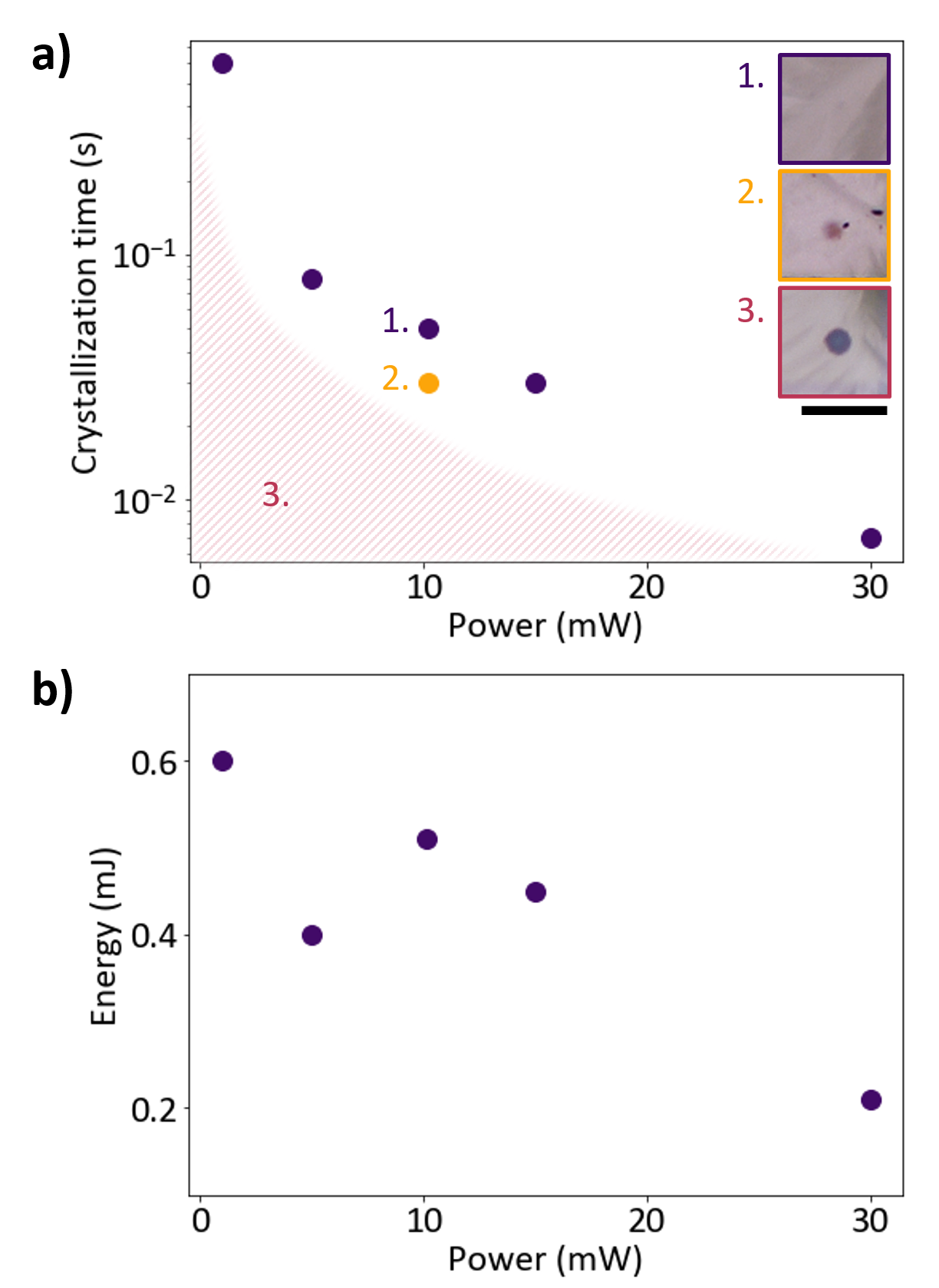}
    \caption{Minimum recrystallization time \textbf{a)} and energy \textbf{b)} as a function of the laser power (in dark purple). The yellow dot in \textbf{a)} is an example of power and time combination to achieve partial recrystallization and the inset represents the microscope views of the sample's surface with a fully-recrystallized (1.), a partially recrystallized (2.) and an initial amorphized (3.) spots for the corresponding data points on the graph (scale bar: 20 µm) - Note that the dots represent the actual experimental values, while the colored dashed part is an estimation of the limits for recrystallisation.}
    \label{Xtal}
\end{figure}

As the amorphization window has been found, the same study can be conducted for the laser induced recrystallization of Sb$_2$S$_3$ in order to find the optimum parameters for the reversible phase transition. The material was first amorphized with the highest possible energy below the damage threshold to guaranty a complete phase transition on the widest possible area (see inset 3. figure \ref{Xtal} \textbf{a}). It was assumed that all amorphized regions made with the same pulse energy are comparable. The amorphized spots are then exposed to the CW laser, with varying exposure time and power. The objective is to find the minimum time/energy set to completely recrystallize the spot. To do so, for a given laser power, the spots were exposed to the CW laser for various exposure times (with a time-step difference of 2 ms) and the surface of the sample was observed under the optical microscope to verify the recrystallization of the spots. A spot was assumed completely recrystallized when the purple amorphized region was no longer visible (see inset 1. figure \ref{Xtal} \textbf{a}). The exposure time was thus increased until the amorphized spots completely disappear and the surface of the sample appears fully crystalline under the optical microscope. This process was conducted for several laser powers.\\

Figure \ref{Xtal} \textbf{a} displays the minimum exposure time needed to achieve the full laser recrystallization of amorphized spots (inset 1.) as a function of the laser power. The results are unambiguous: the higher the power, the faster the recrystallization time. Indeed, the time needed to fully recrystallize the amorphous spot for a 1 mW laser power is 600 ms, while it becomes 7 ms for a 30 mW laser pulse. The most straightforward explanation for this would be energy conservation: considering there is a minimum activation energy to recrystallize the laser-amorphized region, the same total energy can be obtained by compensating a decreased laser power with an increased exposure time. Yet, the experimental values of power and recrystallization time do not yield a constant energy (see figure \ref{Xtal} \textbf{b}), suggesting the existence of an additional factor explaining the observed trend. As we explain in the following, this may be due to the crystallization kinetics of Sb$_2$S$_3$. We showed in a previous work \cite{programPCM} that the crystallization dynamics of Sb$_2$S$_3$ follows the Avrami's law, which states that the crystallization is faster as the temperature increases. This theory takes into account the two steps of crystallization: nucleation and subsequent growth of the grains. Here, the amorphized spots are surrounded by a crystalline matrix, therefore recrystallization just consists in grain growth.
From the crystallization theory  \cite{Wuttig2021thermodynamics}, the grain growth rate of the crystal $u(T)$ as a function of the temperature $T$ is expressed as follows: 
\begin{equation}
    u(T)=f.D(1-e^{\frac{-\Delta G}{R.T}})
\end{equation}
where $f$ accounts for the surface roughness, $D$ is the diffusivity accounting for atomic mobility, $G$ is the thermodynamic driving force and $R$ is the gas constant. From this equation, we see that when the temperature increases, the grain growth is faster. Considering that the temperature inside the Sb$_2$S$_3$ thin film directly increases with higher laser power, the trend in figure \ref{Xtal} can be explained from PCM crystallization kinetics. 

Using these results and the crystallization theory, it is then possible to achieve partial phase transitions of the material by stopping the recrystallization before it is complete, as shown in inset 2. figure \ref{Xtal} \textbf{a}. This partially recrystallized spot was obtained by reducing the exposure time from 50 ms (full recrystallization) to 30 ms for a 10 mW laser pulse (yellow dot in figure \ref{Xtal} \textbf{a}). It is thus possible to program several partially recrystallized states in the depth of the PCM, which is promising for multi-level applications such as grayscale lithography and neuromorphic computing. Additionally, contrary to the laser amorphization, the shape of partially recrystallized spots should not be impacted by the beam profile, as the phase transition is primarily governed by the crystallization dynamics of Sb$_2$S$_3$. Thus this method seems  more promising for controlling the partial phase transition than the usual partial amorphization obtained via increasing the number of pulses. All these results show that the laser amorphization of Sb$_2$S$_3$ is reversible, and, although we did not carry out a full endurance test, the reversible switching could actually be performed over 10 cycles. \\

\section{Discussion}

\subsection{Reproducibility and impact of the material's polycrystallinity}
\begin{figure*}[!ht] 
    \centering
    \includegraphics[width=1\linewidth]{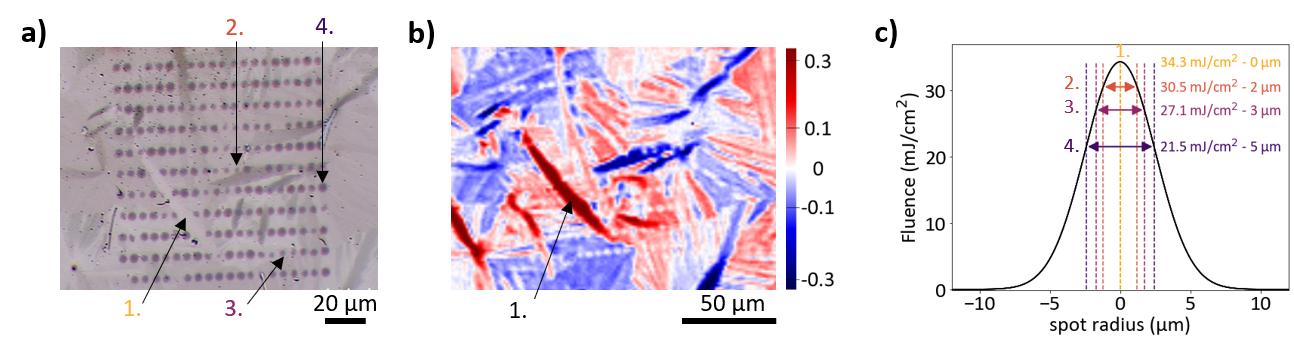}
    \caption{Impact of the anisotropy on amorphization reproducibility shown with \textbf{a)} the optical microscope images and \textbf{b)} the Mueller matrix element M13 of a grid of amorphized spot made with a 13.5 nJ laser pulse on 42 nm-thick Sb$_2$S$_3$ layer. \textbf{c)} Measured diameters of few spots and corresponding fluence threshold  displayed on the calculated Gaussian beam profile for the 13.5 nJ laser pulse.}
    \label{Anisotropy}
\end{figure*}

Exploring the reproducibility of the laser-induced phase transition of Sb$_2$S$_3$ is of major importance for the future integration and scalability of this PCM on devices. To check the reproducibility of our switching method, arrays of 20$\times$11 amorphization spots were made on the 42 nm-thick Sb$_2$S$_3$ for several laser powers. An example of such a grid of spots is given in figure \ref{Anisotropy} for a 13.5 nJ laser pulse. It is clear from the optical microscope view of the sample's surface that the spot sizes are position-dependent (cf. figure \ref{Anisotropy} \textbf{a)}). 
Here, it is important to remember that the thermal crystallization of Sb$_2$S$_3$ results in a randomly oriented polycristalline layer. As Sb$_2$S$_3$ crystallizes in an orthorombic structure, it is anisotropic and its optical properties are therefore orientation-dependent \cite{schubert2004}. Then, due to the polycrystallinity and occurrence of several orientations in the Sb$_2$S$_3$, the layer is also anisotropic with position-dependant properties. This appears clearly on the different microscope images of the sample displayed previously, with large grains of different contrasts, similarly to previous reports \cite{gutierrez2022polarimetry, Yaelinterlaboratory, Liu2020}.
In figure \ref{Anisotropy} \textbf{a)}, we notice a correlation between the size of amorphous spots and crystal grain orientations. This observation indicates that the amorphization threshold is impacted by the crystal grain's orientation on which it is performed. Specifically, amorphization did not occur on the grain located at the lower left part of the spot grid (identified as 1.) in figure \ref{Anisotropy} \textbf{a)}. 

The Mueller matrix element m13 of the same region in figure \ref{Anisotropy} \textbf{b)} clearly highlights the anisotropy of Sb$_2$S$_3$, which is characterized by non-zero values (colored region - see the methods section and supplementary for more information). The grain on which amorphization did not occur appears in dark red, thus showing high anisotropy (see arrow 1. in figure \ref{Anisotropy} \textbf{b)}, while the amorphized spots appear as white and are therefore isotropic. Given the temperature within the Sb$_2$S$_3$ layer is determined by both the thermal properties of the material and its optical absorption profile, the anisotropy of Sb$_2$S$_3$ consequently affects the amorphization threshold through a modification of the optical absorption. So far, the importance of this property of the PCM comparatively to other impacting parameters on the amorphization has not been precisely quantified. By measuring several spot diameters and calculating the corresponding fluence threshold, it appears that the latter fluctuates between 21.5 and at least 34.3 mJ/cm$^2$ (see figure \ref{Anisotropy}  \textbf{c)}, the considered spots are marked on figure \ref{Anisotropy}  \textbf{a)}). The fluence window for amorphization is therefore strongly affected by the anisotropy and polycristallinity of Sb$_2$S$_3$, as it can induce a variation of amorphization threshold up to 36\% of the overall amorphization window. This parameter thus seems to be the most impacting one when performing the optical switch of Sb$_2$S$_3$ and makes it hard to conclude on the impact of other parameters whose variability is typically lower. Although it was not precisely quantified, the polycrystallinity of the film should also have a similar impact on the laser-induced recrystallization. Consequently, the optimal laser power for the reversible amorphization should be calibrated for each grain with a local optical dispersion measurement. 

\subsection{Impact of the film thickness}

So far we have only discussed the reversible laser amorphization of a 42 nm-thick film of Sb$_2$S$_3$. However, for free space nanophotonic applications such as metasufaces, thicker films, typically of several hundred nanometers, are often needed to achieve the desired functionality. Yet, the optical switching of thick PCM films is often reported as a challenging task \cite{zhang2021myths, lawson2024MilionCycle}, thus underlining the necessity to study the impact of the Sb$_2$S$_3$ thickness on the laser amorphization. 

\begin{figure}[ht] 
    \centering
    \includegraphics[width=1\linewidth]{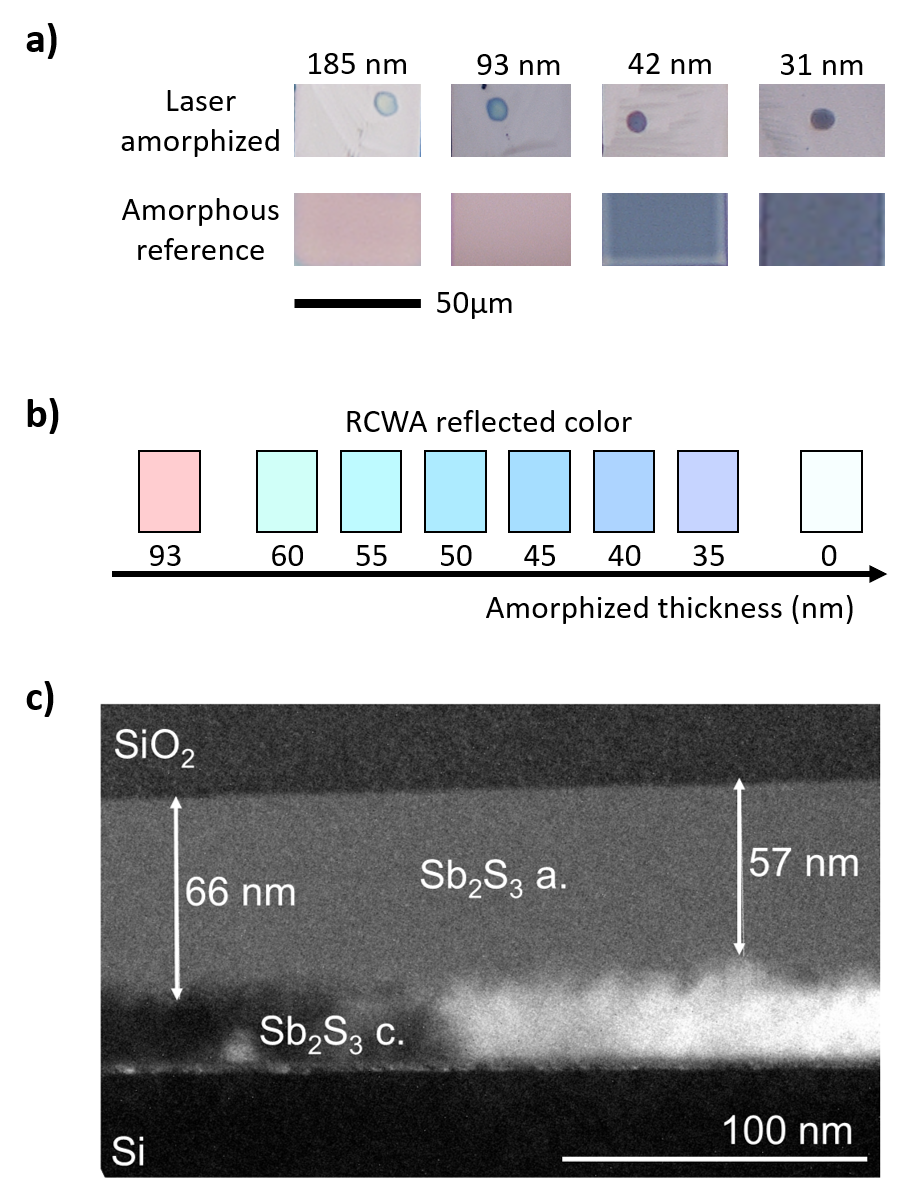}
    \caption{Amorphization of thick films: \textbf{a)} optical microscope image of laser amorphized and as-deposited amorpous sample of different thickness of Sb$_2$S$_3$ \textbf{b)} simulated colors of the stack Si substrate/Sb$_2$S$_3$ c/ Sb$_2$S$_3$ a/ SiO$_2$ (30 nm) for varying amorphized thicknesses on the 93 nm Sb$_2$S$_3$ sample \textbf{c)} TEM cross section of the laser amorphized spot on the 93 nm-thick Sb$_2$S$_3$ sample observed in the dark field mode.}
    \label{thickFilmExp}
\end{figure}

We investigate the laser amorphization of Sb$_2$S$_3$ films with thicknesses ranging from 31 nm to 175 nm. All samples have been fabricated and amorphized using the method presented in the previous sections. Figure \ref{thickFilmExp} \textbf{a)} presents the optical microscope images of spots realized on four Sb$_2$S$_3$ thin films using the maximum laser power before damage. As evidenced by the color change, all samples could be amorphized, regardless of their thickness. The window ranges necessary for amorphization have been determined from the simulations as well as experimentally, as explained in section \ref{Laser amorphization}, and are given in table \ref{TabAmThresThick}. From the simulations, both the amorphization and damage thresholds slightly decrease when increasing the film thickness. This can be attributed to the lower thermal conductivity of Sb$_2$S$_3$ compared to the silicon substrate: the Sb$_2$S$_3$ layer acts as a resistance to the heat dissipation towards the substrate. When increasing Sb$_2$S$_3$ thickness, the heat leakage in the substrate is thus mitigated, hence reducing the energy needed to reach a given temperature. This effect was also reported for Sb$_2$Se$_3$ by Lawson et al. \cite{lawson2024MilionCycle}. Experimentally, this trend does not appear as clearly, except for the damage threshold of the thinner film. 
The fact that all films seem to have more or less the same amorphization and damage thresholds, is attributed to the large uncertainty on the determination of the thresholds. This could be a consequence of the films anisotropy: it was shown previously that the amorphization threshold could vary by nearly 13 mJ/cm$^2$ depending on the position on the Sb$_2$S$_3$ grain. 

\begin{table}[h!]
\centering
\begin{tabularx}{\linewidth}{>{\centering\arraybackslash}m{1cm}|*{2}{>{\centering\arraybackslash}X}|*{2}{>{\centering\arraybackslash}X}}
& \multicolumn{2}{c|}{Sim. fluence (mJ/cm$^2$)} & \multicolumn{2}{c}{Exp. fluence (mJ/cm$^2$)} \\
& Amorphization  T$_{\text{max}}$=T$_{\text{melt}}$ & Damage T$_{\text{max}}$=T$_{\text{evap}}$ & Amorphization & Damage \\
\hline
31 nm & 26 & 71 & 21 - 32 & >95 - <106 \\
42 nm & 22 & 52 & 21 - 29 & >67 - <79 \\
93 nm & 19 & 39 & 18 - 21 & >51 - <68 \\
175 nm & 18 & 40 & 20 - 22 & >65 - <74 \\
\end{tabularx}
\caption{Amorphization and damage fluence threshold in mJ/cm$^2$ extracted from the thermo-optical simulations (sim.) and from the experimental gaussian beam profile (exp.)}
\label{TabAmThresThick}
\end{table}

Nevertheless, increasing the film thickness still induces a difference in the results: the thick films could not be completely amorphized. Indeed, the spots on the Sb$_2$S$_3$ films thicker than 42 nm present different colors than the as-deposited amorphous reference, as shown in figure \ref{thickFilmExp} \textbf{a)}. Additionally, the crystal grains are visible below the spots for these layer thicknesses, suggesting that the films are just partially amorphized. It is possible to estimate the amorphized thickness by colorimetry analysis: we compare the spot color to the simulated reflected color of the stack, as calculated by coupling the reflectivity from transfer matrix with color matching functions (more details in \cite{vilayphone2024design}). An example of such analysis is displayed in figure \ref{thickFilmExp} \textbf{b)} for the 93 nm sample. The simulated reflected colors for the stack Si (substrate)/Sb$_2$S$_3$ c/Sb$_2$S$_3$ a/SiO$_2$(30 nm), are presented in figure \ref{thickFilmExp} \textbf{b)} as a function of the amorphous Sb$_2$S$_3$ thickness (the overall Sb$_2$S$_3$ thickness being 93 nm). From the color of the experimentally measured spot, we estimate an amorphized thickness that lies between 45 nm and 55 nm for this sample. This hypothesis is verified by TEM observations. Indeed, to verify precisely the amorphized thickness, we have conducted a TEM analysis of the spot made on the 93 nm-thick film. Figure \ref{thickFilmExp} \textbf{c)} displays the center of spot cross-section TEM dark-field image, where the crystallinity is illustrated by a strong diffraction contrast and the maximum amorphized thickness is 66 nm in the center of the spot. The amorphous to crystalline interface being rough, the amorphized thickness can only be estimated within $\sim$10 nm precision. The underestimation of the amorphized thickness from the color simulation could be due to this roughness as well as to a potential volume change upon phase transition. These results indicate the existence of a maximum Sb$_2$S$_3$ thickness that can possibly be amorphized in this particular configuration of laser and sample. This phenomenon has also been observed by other groups \cite{lawson2024MilionCycle, gao2024SbSOpticalswitchLambda} and is commonly referred to as the thickness limit \cite{zhang2021myths}. It is essential to overcome this thickness limit in order to optimize the potential of Sb$_2$S$_3$ for various nanophotonic applications.

\subsection{Insights on overcoming the thickness limit}

Successful laser amorphization is the result of a competition between several physical phenomena: the recrystallization kinetics, the thermal properties of the stack and the optical absorption. A commonly given explanation for the thickness limit is the interplay between the thermal conductivity and the critical cooling rate of the PCM \cite{zhang2021myths}. However, we show in this section that the optical absorption is the limiting aspect when amorphizing our thick Sb$_2$S$_3$ films and the maximum amorphized thickness can be increased by optimizing the absorption profile. 

For amorphization to occur, the material must be melt-quenched fast enough to suppress crystallization. The critical cooling rate $\Theta _{crit}$ is defined as the minimum quenching rate preventing crystallization: $\Theta _{crit}=\frac{\Delta T}{\tau_{min}}$ where $\tau_{min}$ is the minimum crystallization time, and $\Delta T$ is the temperature difference between the melting point $T_{melt}$ and the temperature at which crystallization is the fastest $T_{nose}$ \cite{BookVerreXtal}, \cite{Wuttig2021thermodynamics}. It has been reported that for thick films, the PCM thermal conductivity is the limitation to this critical cooling rate as it can hinder the heat evacuation out of the PCM. Considering only the PCM thermal conductivity, the thickness limit $t_{max}$ can be expressed as a function of the critical cooling rate as followed \cite{lawson2024MilionCycle}: 
\begin{equation}
   t_{max}\approx \sqrt{\frac{\alpha .\Delta T}{\Theta _{crit}}} = \sqrt{\alpha .\tau_{min}}
\end{equation}
where $\alpha$, the thermal diffusivity ($\alpha$=$\kappa$/(C$_p$.$\rho$)) is assumed constant within the PCM. For Sb$_2$S$_3$, taking $\tau_{min}=7$ ms, our minimum experimental crystallization time, the calculated thickness limit is 70 µm, which is much larger than the thickness of the films considered in the present work. One could argue that the minimum crystallization can be smaller when using high energy amorphization pulses. Yet, the smallest crystallization time reported is a few µs, which would give a thickness limit around 800 nm. These results could vary a bit with different values of thermal diffusivity but would not change the conclusions presented here. Additionally, our laser pulse duration (500 ps) being much smaller than this minimum crystallization time and the use of Si substrate helping heat evacuation make it very unlikely for Sb$_2$S$_3$ crystallization kinetics to be the reason for thick film partial amorphization. \\

\begin{figure*}[ht] 
    \centering
    \includegraphics[width=1\linewidth]{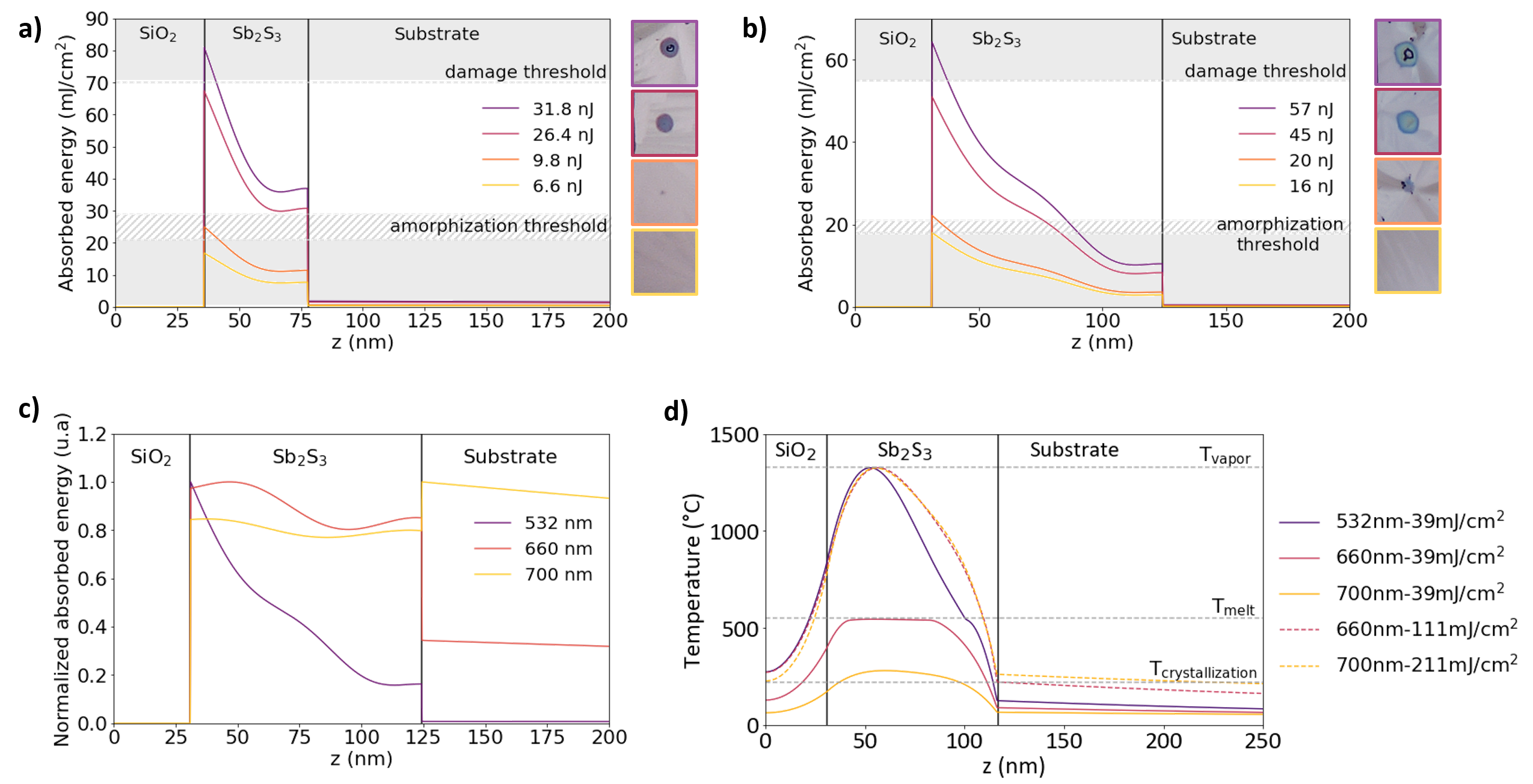}
    \caption{Simulated optical absorption at 532 nm in the center of the spot for different laser energy for the thin film of \textbf{a)} 42nm and  \textbf{b)} 93 nm of Sb$_2$S$_3$ and corresponding top vew of the spots. Impact of laser wavelength on \textbf{c)} the normalized optical absorption profile and \textbf{d)} the temperature profile throughout the sample. }
    \label{ImpactLambda}
\end{figure*}

On the other hand, we explain in the following that the main obstacle for amorphizing thick layers is the strong optical absorption of Sb$_2$S$_3$. Indeed, as explained before, in optically induced phase transitions, the optical absorption directly determines the amount of energy available for the transition to occur. Figure \ref{ImpactLambda} \textbf{a)} and \textbf{b)} display the profile of absorbed energy respectively in the 42 and 93 nm-thick Sb$_2$S$_3$, calculated for several laser powers. When comparing the results to the experimental amorphization window (limits in grey on the graphs) for the two samples, it appears that, at laser energies close to the damage threshold, the absorbed laser power remains above the amorphization threshold throughout the film for the 42 nm-thick film while it goes below the amorphization threshold in the last 37 nm for the 92 nm-thick film. For this latter film, we conclude that it is not possible to fully amorphize it in these conditions, as the absorbed energy decreases too sharply within the film. Thus, the absorbed energy density in the last 37 nm ends up below the amorphization threshold when the center of the spot starts to be damaged. One way to overcome this optical thickness limit would be to flatten the absorption profile. This could be realized by shifting to laser wavelengths that are less absorbed by Sb$_2$S$_3$. Figure \ref{ImpactLambda} \textbf{c)} shows that by increasing the laser wavelength up to 660 nm, the absorbed energy profile is smoothed out. The thermo-optical simulations in figure \ref{ImpactLambda} \textbf{d)} demonstrate that the resulting temperature profile also flattens when increasing the laser wavelength. It results in an amorphized thickness increasing from 69 nm (wavelength of 532 nm) to 79 nm (wavelengths of 660 nm and 700 nm), by scaling up the laser energy to have the same maximal temperature for all laser wavelengths. Recently, Gao et al. demonstrated experimentally that 70 nm-thick Sb$_2$S$_3$ on glass could not be amorphized completely with a 480 nm laser, while amorphization was reported to be total when using a 550 nm laser \cite{gao2024SbSOpticalswitchLambda}, which corroborates our theory. The downside of this method is that more energy is absorbed in the substrate due to the lower imaginary part of Sb$_2$S$_3$ refractive index at these wavelengths as evidenced in figure \ref{ImpactLambda} \textbf{c)}. This consideration explains the drastic increase of laser fluence necessary to reach a given temperature. Indeed, from the thermo-optical simulation the laser energy density had to be increased from 39 mJ/cm$^2$ to 111 mJ/cm$^2$ and 211 mJ/cm$^2$, respectively at 660 nm and 700 nm, for the maximum temperature in the PCM to be the same as with the 532 nm pulse. 

This analysis revealing that the bottleneck for the thickness limit of amorphization is not due to intrinsic thermal properties of Sb$_2$S$_3$ is actually encouraging. Indeed, increasing the thickness limit of amorphization is therefore a matter of engineering the optical and thermal environment of Sb$_2$S$_3$ so as to optimize the absorption and heat distribution within the PCM layer.

\section{Conclusion}

The reversible laser amorphization of Sb$_2$S$_3$ layers of various thicknesses was successfully demonstrated. The experimental results were related to multi-physics simulations for a better understanding of the underlying phenomena involved in the optically induced phase transition of Sb$_2$S$_3$. The experimental amorphization fluence threshold was found to be between 18 mJ/cm$^2$ and 29 mJ/cm$^2$, while the films started to be irreversibly damaged for laser energy density above 51 mJ/cm$^2$. Overall, the thermo-optical simulations were in good agreement with these results, validating the proposed theoretical model. There was no significant impact of the thin film thickness on the amorphization energy window. However, films thicker than 42 nm could not be completely amorphized contrarily to the thinner films. This limitation was attributed to the sharpness of the optical absorption profile in the depth of the PCM and could be overcome by increasing the laser wavelength up to 660 nm. Additionally, we report the strong impact of the polycristallinity and anisotropy of Sb$_2$S$_3$ on the amorphization results. This polycristallinity induced a 36\% variation on the value of the amorphization threshold, which is higher than the variability induced by the film thickness. Controlling the local optical absorption is thus of major importance for optimizing Sb$_2$S$_3$ laser amorphization since it was found to be the most impactful parameter. Regarding the recrystallization of the spots, full recrystallization was performed with a minimum time of 7 ms for a 30 mW laser power. The recrystallization time decreased exponentially when increasing the laser power, suggesting that the minimum recrystallization time could be further lowered by using a laser with a higher power. Moreover, not only did we achieve the complete recrystallization of the amorphous spot, but also the partial laser-induced recrystallization of Sb$_2$S$_3$, which is a promising way to obtain multi-level phase transition in the depth of Sb$_2$S$_3$. Therefore, our work paves the way for a full control of the all-optical reversible switching of Sb$_2$S$_3$ and shows promise for rewritable patterns, encoding information in the depth of PCM and designing dynamically tunable images, holograms and devices.

\section{Methods}

\subsection{Sample fabrication}

The Sb$_2$S$_3$ and SiO$_2$ layer were deposited on silicon substrate by electron beam evaporation at a rate between 0.6 and 1 Å/s and at an initial pressure around 10$^{-6}$ mbar. The Sb$_2$S$_3$ is deposited amorphous and subsequently crystallized thermally using a linkam heat cell set at 280°C. The crystallization is realized under ambient pressure. The laser amorphization is performed on the thermally crystallized samples.

\subsection{Determination of the amorphization threshold}

To calculate the amorphization threshold, we measured the average power at the output of the optical setup where we position the sample using a powermeter. We then calculate the energy from the laser repetition rate and input it in the Gaussian beam formula. The spot diameters are measured under the optical microscope with the microscope software, with a precision of 0.5 µm approximately. The amorphization threshold is the calculated fluence from the Gaussian beam formula for a given energy E and the corresponding spot radius r.

\subsection{Simulations}

The multi-physics simulations were carried out using MATLAB 2023a.
The optical stage is a re-implementation of the transfer-method matrix \cite{mackay2022transfer}, while the thermal stage is carried out using MATLAB's partial differential equation solver.
Given an initial multilayer structure, the optical stage is run first, and the calculated absorbed power density is used as the heat source term in the thermal stage.
The thermal simulation is interrupted as soon as 1 nm of crystalline material overshoots the melting point.
When this occurs, the multilayer structure is updated to reflect the change in material properties, and both simulation stages are run again.
Improper discretization of the problem can be a large source of numerical error, especially due to interfaces where thermal parameters are discontinuous.
Hence, a sub-picometric meshing is enforced around all layer interfaces, and the meshing is updated in step with the structure throughout the duration of the simulation.

The main material parameters used are summarized in table \ref{tab:sim-params}. The initial and back plate temperatures were both set to 20°C.

\begin{table}[h]
\begin{tabular}{c|ccccc}
 & n  & k   & $\mathrm{C_p}$  & $\mathrm{\rho}$      & $\mathrm{\kappa}$   \\
 &  &   & J.K$^{-1}$.kg$^{-1}$ & kg.m$^{-3}$     & W.K$^{-1}$.m$^{-1}$   \\
\hline
$\mathrm{Sb_2S_3}$ a   & $\mathrm{3.55 \pm 0.01}$  & $\mathrm{0.18 \pm 0.03}$  & 353 \cite{HemmatyarMetaDisplay}  &    4600 \cite{Liu2020}       & 1.16 \cite{Liu2020}   \\
$\mathrm{Sb_2S_3}$ c & $\mathrm{4.72 \pm 0.03}$ & $\mathrm{0.74 \pm 0.04}$ &    353 \cite{HemmatyarMetaDisplay}   & 4600 \cite{Liu2020} &   1.16 \cite{Liu2020} \\
SiO2              & 1.47 &   0  &   709  & 2203  & 1.38     \\
Si                &  4.15  &  0.03
& 711 & 2330      & 148

\end{tabular}

\caption{\label{tab:sim-params}Material parameters used for the simulations}
\end{table}

The n and k values are experimental values taken from ellipsometry measurements while the thermal parameters are taken from the literature ($\mathrm{Sb_2S_3}$) or Lumerical database ($\mathrm{SiO_2}$ and Si). The latent heat, representing the energy required to melt the PCM, is assumed to be $\mathrm{\Delta H_f \approx 120  kJ/kg}$ \cite{johnson1981enthalpies}.


\subsection{TEM characterization}

Conventional transmission electron microscopy (TEM) observations were performed using a JEOL 2100. The thin foils were prepared from the selected samples via the lift-out method using a Ga LMIS focused ion beam instrument (Zeiss Crossbeam 550L). The TEM was operated at 200 kV and is equipped with a high tilt objective lens pole piece. Images were recorded using a bottom mounted Gatan Orius SC1000 CCD camera with a physical pixel size of 9$\times$9 µm. Care was taken to limit the electron dose on the observed foils in order to avoid thermally induced crystallisation of the amorphous Sb$_2$S$_3$ layer by the electron beam. Bright field and dark field images were obtained by placing a 20 µm aperture near the back focal plane of the objective lens; this corresponds to a collection angle across the aperture disk of 4 mrad. The dark field images were obtained by tilting a diffraction spot of the crystalline Sb$_2$S$_3$ onto the optical axis of the microscope. Amorphous parts of the thin foil are visible via mass thickness scattering, which gives a weaker contrast with respect to the diffraction contrast of the c-Sb$_2$S$_3$.

\subsection{Mueller matrix characterisation}

In this work, we report the microscopic maps of the Block off diagonal element of the normalized Mueller matrix recorded with a imaging spectroscopic ellipsometer (Accurion EP4, Park Systems GmBh, G\H{o}ttigen Germany) equipped with a 20x objective at a 420 nm wavelength, and an angle of incidence of 50°. We only displayed the normalized m13 element, accounting for diattenuation (preferential attenuation of one polarization relatively to its orthogonal counterpart) of linearly polarised light along the -45°/+45°. This element was chosen as it evidences best the position-dependant properties of crystalline Sb$_2$S$_3$. However, for a complete understanding of the material properties the full Mueller matrix is needed. The latter can be found in the supplementary material.

\begin{acknowledgments}
We acknowledge funding from the French National Research Agency (ANR) under the projects MetaOnDemand (ANR-20-CE24-0013) and OCTANE (ANR-20-CE24-0019).
SVM and SE acknowledge the project ADONIS (No. CZ.02.1.01/0.0/0.0/
16-019/0000789) from the European Regional Development Fund. Y.G. acknowledges founding from a Ramon y Cajal Fellowship (RYC2022-037828-I). 
We thank F. Bentata, H. S. Nguyen and X. Letartre for fruitful discussions, A. Benamrouche for AFM images and D. Troadec for the FIB preparation.

\end{acknowledgments}

\medskip

\bigskip



\nocite{*}
\bibliography{TestBiblio}

\end{document}